%% file: manuscript.tex
\documentclass[10pt,journal]{IEEEtran}
\usepackage[utf8]{inputenc}
\usepackage{amsmath,amssymb,graphicx,booktabs,cite,url,algorithmic}
\usepackage[hidelinks]{hyperref}
\graphicspath{{figures/}}
\newcommand{\W}{\mathcal W}

\newcommand{\NT}{N_{\mathcal T}}
\begin{document}
\bstctlcite{paper5bibcontrol}
\title{Tile-Based Phase Unwrapping with Reliability-Ordered Joining and Residual-Weighted Multipath Averaging}
\author{Antoine~Moevus and Max~Mignotte\thanks{The authors are with the D\'epartement d'informatique et de recherche op\'erationnelle, Universit\'e de Montr\'eal, Montr\'eal, QC, Canada. E-mail: antoine.moevus@umontreal.ca; max.mignotte@umontreal.ca.}}
\maketitle
\begin{abstract}
Phase unwrapping recovers a continuous phase map from measurements known only modulo $2\pi$, in the presence of noise and discontinuities. We propose a tile-based method that combines local spectral reconstruction, reliability-ordered tile joining, and residual-weighted multipath averaging. Each tile is reconstructed by a discrete cosine transform (DCT) least-squares solve, and robust measurements along tile borders guide their joining along a spanning tree. We repeat this reconstruction on shifted tile grids and on rotated and reflected copies of the image, which changes tile borders and joining paths. A residual-weighted average combines the resulting reconstructions, which are computed in parallel and averaged on the fly without being stored. We also introduce the corrected circular rewrapping error, a congruence measure: after removing a global phase offset, it measures how far the rewrapped reconstruction is from the observed phase. It applies to least-squares and path-based methods, even without an unwrapped reference. Used with reference-based errors, it separates agreement with the measurement from recovery of the correct number of cycles. Experiments on radar-interferometric, magnetic-resonance, holographic, and constructed phase images compare our method with seven reference algorithms. Among the methods that complete every test image, ours has the lowest mean fraction of pixels with an error above $\pi$. On the images all methods complete, it lies on several speed--accuracy Pareto fronts. Error maps also reveal isolated cycle errors that average scores or congruence alone can hide.
\end{abstract}

\section{Introduction}
\IEEEPARstart{P}{hase} measurements in interferometry and magnetic resonance are known only modulo a full cycle. The wrapped observation alone does not tell how many full cycles to add at each pixel, so unwrapping needs assumptions about how the phase varies in space, about the noise, or about the acquisition \cite{itoh1982analysis,ghiglia1998two}. Discontinuities, undersampling, and noise can make the differences between neighboring pixels inconsistent. Each unwrapping method then decides where this inconsistency goes in the result.

Two families of methods handle this inconsistency differently. Reliability-guided integration chooses which local differences to follow, whereas a global least-squares (Poisson) solution spreads the mismatch over the whole image \cite{arevalilloherraez2002fast,ghiglia1994robust}. Local least-squares solves with the discrete cosine transform (DCT) lie in between: each solve covers only a small tile, but leaves one unknown constant per tile. Finding these constants is a second integration problem. Moreover, with a single tile grid, the result depends on where the tile borders fall.

We propose a tile-based method with three steps: a local DCT reconstruction of each tile, a joining of the tiles in order of reliability, and a weighted average of several reconstructions. Measurements along each tile border give the offset between two neighboring tiles; the spread of these measurements, their median absolute deviation, sets the order in which a spanning tree connects the tiles. The whole solve-and-join procedure is repeated on shifted grids and on rotated or reflected copies of the image. We call each resulting full-image reconstruction a pass, and the rotations and reflections image isometries. A residual score based on the Laplacian sets the weight of each pass in the final average.

The method builds on known ideas: block least-squares reconstruction, reliability-guided merging, and cycle spinning \cite{strand1999two,strand2002two,antonopoulos2015tilebased,coifman1995translationinvariant}. Chen et al.\ also combine blockwise DCT least-squares unwrapping with stitching-path optimization \cite{chen2025stitching}. Our contribution is the way four parts work together: border measurements that use the wrapped phase difference across the border, a robust ranking of the borders, real-valued tile offsets, and a residual-weighted average of shifted reconstructions. We evaluate the complete method on several imaging modalities.

Our second contribution concerns evaluation. Least-squares solvers, global or tiled, fit the phase gradients; once rewrapped, their output need not equal the measured phase. Path-integration and integer-cycle methods, in contrast, keep this equality, called congruence \cite{ghiglia1998two,chen2001twodimensional}. To measure how far a reconstruction is from congruence, we introduce the corrected circular rewrapping error, $\mathrm{MAE}_{\mathrm{uw2}}$ (Section~\ref{sec:maeuw2}). It compares the reconstruction with the observation on the phase circle, so that values close to $-\pi$ and $+\pi$ are not counted as far apart, and it removes an arbitrary global phase offset. It applies to congruent and noncongruent methods alike, even without an unwrapped reference. It complements reference-based errors: it measures agreement with the observation modulo one cycle, not whether the number of cycles is correct.

We compare the method with seven reference algorithms on images from several modalities. Phase errors, structural similarity, and threshold fractions measure different aspects of quality. Error maps, such as the constructed Tree example in Figure~\ref{fig:tree-regions}, show the difference between isolated cycle errors and errors spread over the image. Section~\ref{sec:method} presents the reconstruction and its parallel implementation; Section~\ref{sec:evaluation-measures} distinguishes reconstruction accuracy from observation congruence. The protocol and results follow in Sections~\ref{sec:protocol} and~\ref{sec:results}.

\section{Related Work}
\label{sec:related}
\subsection{Local least-squares reconstruction and tile joining}
Global least-squares unwrapping integrates wrapped differences through a Poisson equation \cite{ghiglia1994robust,ghiglia1998two}. It does not depend on an integration path, but it spreads the effect of inconsistent gradients over the whole image. Block methods solve smaller local problems, then estimate the offsets between blocks \cite{strand1999two}. Robust derivative estimation and adaptive integration provide another way to combine local information \cite{strand2002two}. Ching et al.\ already unwrapped separate regions and joined them in the order of a minimum spanning tree, built with Kruskal's algorithm, whose weights are the widths of the gaps between regions \cite{ching1992two}. Our joining follows the same tree rule, with weights computed from the agreement of border measurements. These works show that the size of the local problem, the estimation of derivatives, and the integration order are separate design choices.

Antonopoulos et al.\ make this separation explicit in a modular tile framework \cite{antonopoulos2015tilebased}. Their model-based tile solver fits polynomial derivatives and restores congruence with the observed phase. Their merger estimates integer-cycle shifts from mean border differences. For reliability-guided merging, a tile's reliability is the inverse mean variance of its junctions, and a junction's reliability is the product of the two tile reliabilities. The DigiHolo software of Antonopoulos et al.\ \cite{antonopoulos2015tilebased} supplies the corresponding tile-based reference implementation in our comparisons. Their reliability computed from tile borders, and their merging in a non-continuous order, are direct antecedents of our method.

Chen et al.\ combine blockwise DCT least-squares reconstruction with region-diffusion stitching for rough-surface interferometry; their path planning uses quality analysis of the wrapped phase blocks \cite{chen2025stitching}. Their work shows that block quality and the choice of stitching path matter as much as the local solve.

\subsection{Spatial integration and network methods}
Sorting by reliability following a noncontinuous path (SRNCP) selects a pixel-level integration order by reliability \cite{arevalilloherraez2002fast}; the statistical-cost network-flow algorithm for phase unwrapping (SNAPHU) reconciles phase differences through a global optimization \cite{chen2001twodimensional,chen2002phase}. ROMEO constructs a reliability-weighted minimum spanning tree for phase integration \cite{dymerska2021phase}. Branch-cut methods identify residues, where wrapped differences around an elementary pixel loop have a nonzero sum, and restrict integration across cuts connecting those residues \cite{goldstein1988satellite}. Achard-de Lustrac et al.\ formulate phase unwrapping and gradient correction on a Delaunay graph, including an $L^1$ minimum-cost-flow solver \cite{achardelustrac2025l1norm}. Together, these methods cover path integration, tile merging, and global optimization of the gradients, which makes them useful points of comparison for our local spectral approach. In our comparison, every method receives only the wrapped phase. This gives all methods the same input, but leaves aside information that applications may use, such as magnitude, coherence, or acquisition parameters.

\subsection{Reconstruction diversity and cycle spinning}
Cycle spinning averages results computed on shifted copies of the data, so that the result depends less on where a fixed grid falls \cite{coifman1995translationinvariant}. Here, shifting the grid changes which pixels lie near tile borders and which border measurements are available. It does not make each pass accurate, and errors shared by all passes survive the average. Rotations and reflections of the image add a second source of variation. For example, when two borders have exactly the same weight, the tie is broken by a fixed enumeration order, and this order changes when the image is rotated; numerical rounding changes as well.

In summary, our method combines three elements: a local DCT reconstruction, without forcing the result back to congruence at the end; real-valued tile offsets estimated with the wrapped phase difference across each border; and a spread score computed directly for each border. It repeats this reconstruction over shifted grids and image isometries, which changes the tile borders and the joining paths, and averages the results, so that the outcome depends less on any single grid or joining order. The comparisons in Section~\ref{sec:results} evaluate the complete method against global spectral, pixel-integration, network-flow, and tile-based reference algorithms.

\section{Proposed Method}
\label{sec:method}
Local DCT solves first reconstruct each tile. Measurements on the tile borders then give the offsets between neighboring tiles, and a tree joins the tiles. Repeating these steps for shifted grids and image isometries gives full-image passes, which are aligned and combined by a residual-weighted average. We define these steps first, then describe their parallel implementation.

\subsection{Observation model and local DCT reconstruction}
\label{sec:local}
Let $\psi=\W(\phi+\nu)$ denote a wrapped observation on a rectangular grid, where $\phi$ is the underlying phase and $\nu$ is the measurement noise, when such a model applies. We use $\W(t)=t-2\pi\lfloor(t+\pi)/(2\pi)\rfloor$, with range $[-\pi,\pi)$, and work in radians. Endpoint conventions in the numerical implementation are specified in Appendix~\ref{app:implementation}. Adding $2\pi n$ to $\phi$, for any integer-valued field $n$, gives the same observation; an assumption on how the phase varies in space is needed to remove this ambiguity. Under the Itoh condition $|\Delta\phi|<\pi$ on every neighboring pair and in the absence of noise, $\Delta\phi=\W(\Delta\psi)$ \cite{itoh1982analysis}.

For one tile, let $g_x=\W(\Delta_x\psi)$ and $g_y=\W(\Delta_y\psi)$ on its internal horizontal and vertical pixel pairs. The local reconstruction minimizes
\begin{equation}
 u=\mathop{\arg\min}_{f}\ \|\Delta_x f-g_x\|_2^2+\|\Delta_y f-g_y\|_2^2.
 \label{eq:local-ls}
\end{equation}
Gradients use forward differences and the divergence uses backward differences, its adjoint. With zero flux at the tile boundary, the normal equations are $Lu=\rho$, where $L=\operatorname{div}\nabla$ is the standard five-point Laplacian and $\rho=\operatorname{div}g$. For a tile of side $\NT$, the DCT diagonalizes $L$:
\begin{equation}
 \widehat u_{r,s}=\frac{\widehat\rho_{r,s}}
 {2\cos(\pi r/\NT)+2\cos(\pi s/\NT)-4},\quad(r,s)\ne(0,0).
 \label{eq:dct}
\end{equation}
Here $r,s\in\{0,\ldots,\NT-1\}$ index the DCT coefficients. The zero-frequency (DC) coefficient is set to zero, so each tile solution has zero mean, and an inverse DCT gives $u$. Equation~\eqref{eq:local-ls} fits the gradients, not the phase values: the rewrapped solution $\W(u)$ need not equal $\psi$.

No spectral filtering (shrinkage) is applied to the DCT coefficients. When the measured gradient field is integrable, the local solve recovers a phase with exactly these gradients, up to a constant, including any noise that is itself integrable. What remains is to find the constant of each tile from measurements on the shared borders.

\subsection{Border offset estimation}
\label{sec:border-offsets}
\begin{figure*}[!t]
\centering\includegraphics[width=\textwidth]{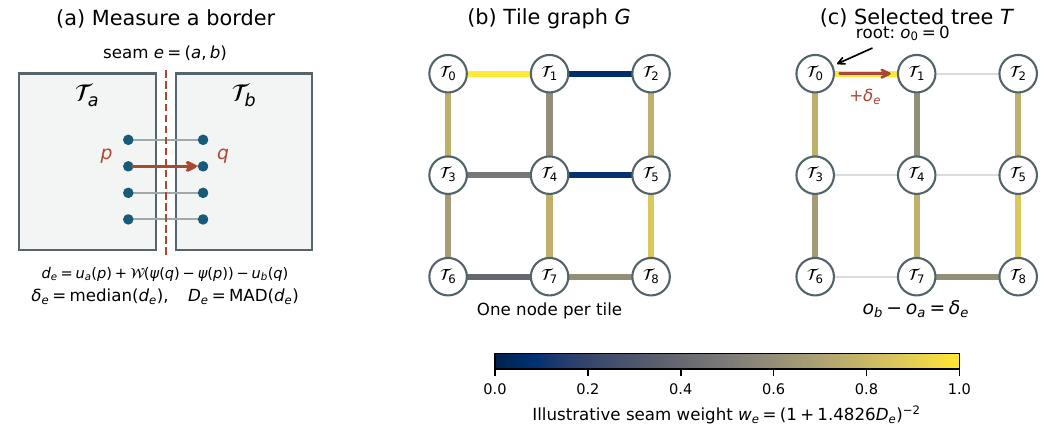}
\caption{A seam measurement becomes a tile-offset constraint. (a) Adjacent pixels $p\in\mathcal T_a$ and $q\in\mathcal T_b$ supply $d_e$ in \eqref{eq:sample}; the median estimates $\delta_e$, and the median absolute deviation (MAD) determines the ranking weight $w_e$. (b) Each tile is a node and each shared border is an edge. (c) Kruskal retains a spanning tree; offsets propagate with the signed rule $o_b-o_a=\delta_e$. Panels (b) and (c) use the same illustrative weights, expressed with $c=1$ and 256 units per cycle.}
\label{fig:seam-tree}
\end{figure*}

Each tile is reconstructed with its own arbitrary phase offset. We call the border shared by two tiles a seam. To assemble a single image, we estimate relative offsets from pixel pairs across shared borders, then propagate them along a tree chosen from the most consistent border measurements (Fig.~\ref{fig:seam-tree}).

For each tile $\mathcal T_a$, the local DCT solver provides a phase field $u_a$. Joining the tiles requires an additive offset $o_a$ for each field:
\begin{equation}
 v(p)=u_a(p)+o_a,\qquad p\in\mathcal T_a.
\end{equation}
Orient each border $e=(a,b)$ from left to right or from top to bottom. The set $B_e$ contains adjacent pixel pairs $(p,q)$ with $p\in\mathcal T_a$ and $q\in\mathcal T_b$. To reproduce the observed wrapped difference at such a pair, the offsets would need to satisfy
\begin{equation}
 \begin{split}
 o_b-o_a&=d_e(p,q),\\
 d_e(p,q)&=u_a(p)+\W(\psi(q)-\psi(p))-u_b(q).
 \end{split}
 \label{eq:sample}
\end{equation}
The wrapped phase difference across the border removes the true phase variation from this offset measurement. In the noiseless case $\psi=\W(\phi)$, if the local solutions satisfy $u_a=\phi+c_a$ and $u_b=\phi+c_b$, and the cross-border differences satisfy the Itoh condition, then $d_e=c_a-c_b$ everywhere on that seam. Its dispersion is zero even when the true phase gradient varies along the border. By contrast, the raw difference $u_a(p)-u_b(q)$ still contains that variation. Noise, undersampling, and imperfect local solutions make these measurements disagree. Their median gives the relative offset $\delta_e$, and their median absolute deviation (MAD) gives their spread $D_e$:
\begin{align}
 \delta_e&=\operatorname{median}_{B_e}d_e,\label{eq:delta}\\
 D_e&=\operatorname{median}_{B_e}|d_e-\delta_e|.\label{eq:mad}
\end{align}
A small $D_e$ means that the measurements agree; however, a seam whose measurements are all wrong by the same full cycle also has a small $D_e$. For an even sample count, each median is the arithmetic mean of the two central ordered values.

We assign each border the weight
\begin{equation}
 w_e=\left[\max(1.4826D_e+c,\epsilon)\right]^{-2}.
 \label{eq:weight}
\end{equation}
The constant $c=2\pi/256$ radians, one encoded phase level, limits the weight contrast between seams, and $\epsilon=10^{-3}(2\pi/256)$ radians is a numerical floor. A seam with a smaller spread receives a larger weight, and the weights set the order in which seams enter the tree.

\subsection{Reliability-ordered tree joining}
The tile graph $G=(V,E)$ has a node for each tile and an edge for each seam. Kruskal's procedure sorts the edges by decreasing weight $w_e$ and keeps an edge only if it connects two groups of tiles that are not yet connected. Ties between equal weights are broken by a fixed order: tiles row by row, right border before lower border. The resulting maximum-weight spanning tree $T$ contains $|V|-1$ edges when the graph is connected.

Set one root offset to zero and traverse the tree, imposing
\begin{equation}
 o_b-o_a=\delta_e,\qquad e\in T.
 \label{eq:propagation}
\end{equation}
When the tree is traversed against an edge's orientation, the sign changes. The resulting offsets satisfy every seam of the tree exactly, before the final rounding of the pixels; seams outside the tree can remain inconsistent. A weak seam may still be needed to connect part of the image, and one wrong seam in the tree shifts every tile behind it.

Although it uses local weights $w_e$, the joining solves a global problem over all tiles. Write the seam residual of an offset field $o=\{o_a\}$ as $r_e(o)=o_b-o_a-\delta_e$. The propagated offsets minimize the weighted count of violated seam constraints,
\begin{equation}
 E_0(o)=\sum_{e\in E}w_e\,\mathbf 1\!\left[r_e(o)\neq0\right],
 \label{eq:E0}
\end{equation}
provided that, around every cycle of the tile graph, the seam offsets $\delta_e$, with their signs, do not sum exactly to zero; this holds for real-valued medians computed from noisy data. Under this condition, the seams satisfied by any offset field contain no cycle and therefore form a forest. Minimizing $E_0$ amounts to maximizing the total weight of a satisfied forest. With positive weights, this maximum is a maximum-weight spanning tree, which Kruskal's greedy rule returns \cite{kruskal1956shortest}, and \eqref{eq:propagation} satisfies every one of its seams. The tree thus trusts the most reliable constraints exactly and ignores the others, whereas a least-squares reconciliation, minimizing $\sum_e w_e r_e(o)^2$, would let every inconsistent seam pull on its neighbors. Minimizing $E_0$ does not minimize the final root-mean-square phase error.

Algorithm~1 summarizes the joining. It is a single direct pass: no iterative solver, such as Jacobi or Gauss--Seidel, is involved, and each local solve is an exact DCT inversion.

\smallskip\noindent\begin{minipage}{\columnwidth}
\hrule\smallskip
\textbf{Algorithm 1:} Reliability-ordered tree joining of one pass.
\smallskip\hrule\smallskip
\begin{algorithmic}[1]
\REQUIRE seams $e=(a,b)$ with offsets $\delta_e$ and weights $w_e$
\STATE sort the seams by decreasing $w_e$; break ties by the fixed enumeration
\STATE $T\leftarrow\emptyset$; each tile forms its own component
\FOR{each seam $e=(a,b)$ in this order}
\IF{$a$ and $b$ lie in different components}
\STATE $T\leftarrow T\cup\{e\}$; merge the two components
\ENDIF
\ENDFOR
\STATE $o_{\mathrm{root}}\leftarrow0$; traverse $T$, setting $o_b\leftarrow o_a+\delta_e$, with the sign reversed against the seam orientation
\RETURN the tile offsets $o$
\end{algorithmic}
\smallskip\hrule
\end{minipage}
\smallskip

In Figure~\ref{fig:seam-tree}(a), $p$ and $q$ are the samples in \eqref{eq:sample}. Panel (b) represents the seams as weighted graph edges. Panel (c) shows which constraints enter \eqref{eq:propagation}. The figure's weights are illustrative. The graph covers the whole rectangular working image; tiles outside the evaluation mask are not removed from it.

\subsection{Grid shifts and image isometries}
\begin{figure*}[!t]\centering
\includegraphics[width=\textwidth]{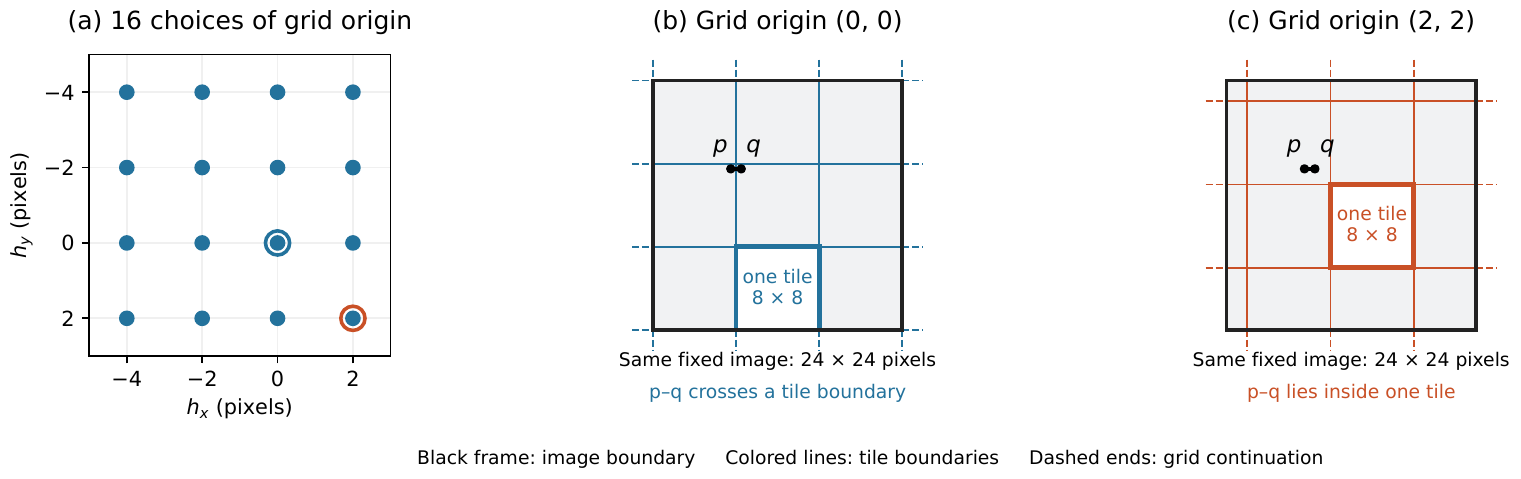}
\caption{Moving an $8\times8$ tile grid over a fixed image. (a) Four horizontal offsets and four vertical offsets, spaced two pixels apart, give sixteen placements of the grid. Each point $(h_x,h_y)$ specifies a shift of the entire tile grid by $h_x$ pixels horizontally and $h_y$ pixels vertically: the tile boundaries move while the image stays fixed. The circled points select the placements illustrated in (b) and (c). (b) A schematic $24\times24$-pixel image, outlined in black, is partitioned at origin $(0,0)$. Colored lines delimit tiles, one of which is highlighted; dashed ends show that the grid continues. (c) The same image and sample pair after moving the grid two pixels right and down. The pair $p,q$ now belongs to one tile instead of straddling a tile boundary. The image size is chosen only for illustration; the tile size and stride match the evaluated configuration.}\label{fig:grid-isometries}
\end{figure*}

Let $\mathcal P_h$ denote the local solve and join for grid origin $h$. For even $\NT$ and stride $s$ dividing $\NT$, the implementation uses origins in
\begin{equation}
 \mathcal H_s=\{-\NT/2,-\NT/2+s,\ldots,\NT/2-s\}^2.
\end{equation}
A pass $k=(h_k,\ell_k)$ uses one grid origin and one square isometry $\mathcal I_{\ell_k}$:
\begin{equation}
 v_k=\mathcal I_{\ell_k}^{-1}\mathcal P_{h_k}(\mathcal I_{\ell_k}\psi).
\end{equation}
The grid moves over a fixed image; the image itself is not translated. Figure~\ref{fig:grid-isometries} shows the grid origins and the resulting tilings. In panels (b) and (c), the same adjacent pixels $p$ and $q$ lie on opposite sides of a tile boundary for one placement and inside a single tile for the other. Their wrapped difference therefore contributes to joining in the first pass and to a local least-squares solve in the second. Combining placements reduces dependence on a particular division into tiles.

\begin{figure*}[!t]\centering
\includegraphics[width=\textwidth]{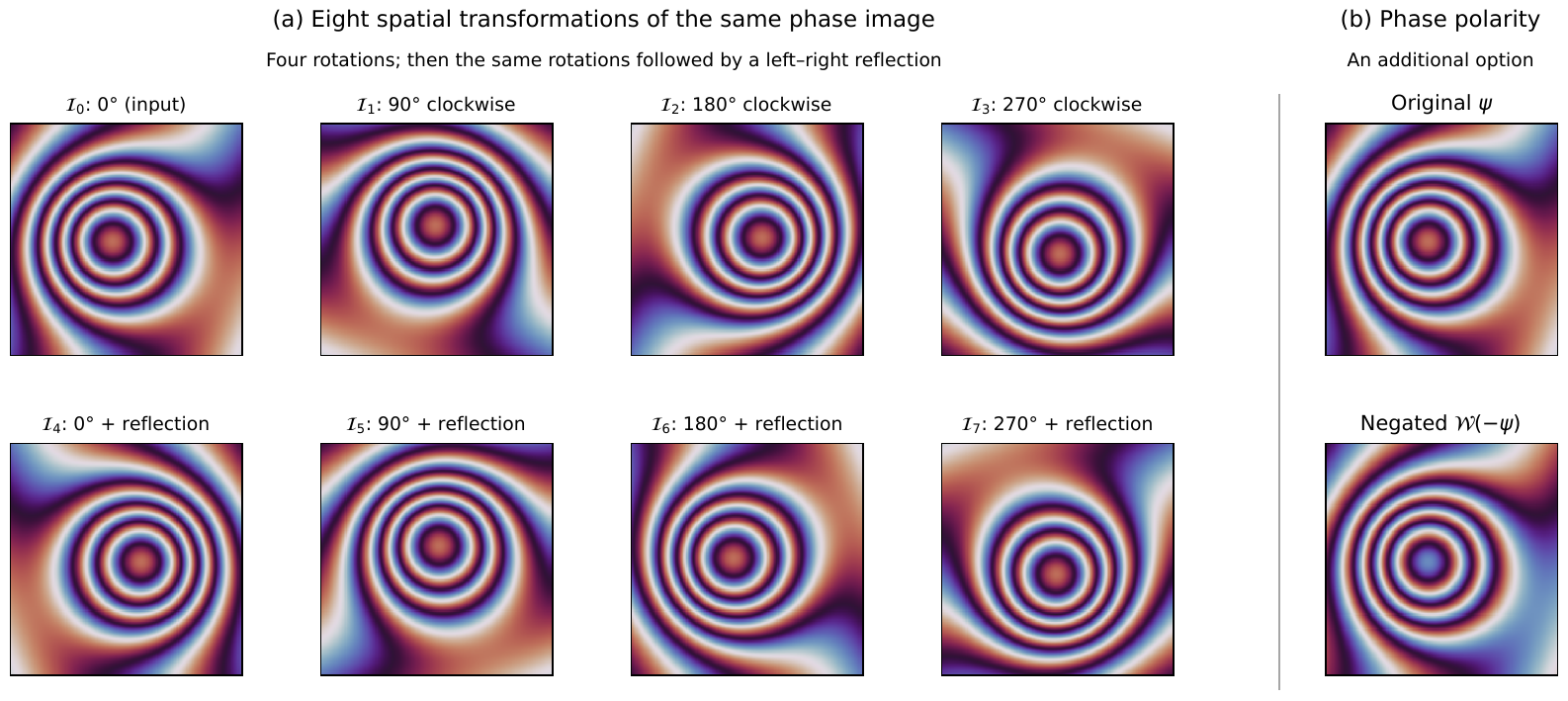}
\caption{Spatial transformations and phase-polarity reversal illustrated on a synthetic wrapped phase image. (a) The upper row shows clockwise rotations through $0^\circ$, $90^\circ$, $180^\circ$, and $270^\circ$; the lower row applies a left--right reflection after each corresponding rotation. The transformed reconstruction is mapped back to the original pixel coordinates before averaging. These eight transformations, each combined with the sixteen grid placements in Figure~\ref{fig:grid-isometries}, form the 128-pass schedule. (b) Negation changes the phase values while leaving pixel positions unchanged. This additional polarity option is shown for clarity but is not included in that schedule.}\label{fig:phase-transformations}
\end{figure*}

The eight square isometries are illustrated with phase images in Figure~\ref{fig:phase-transformations}(a). They comprise four rotations and their left--right reflections. Unlike a grid shift, an isometry transforms the entire input image. Each pass is reconstructed in the transformed coordinates and mapped back by $\mathcal I_{\ell_k}^{-1}$, so corresponding pixels can be averaged. They change the orientation of the image seen by the local solve and by the joining. For the evaluated tile side $\NT=8$, the main schedule combines all eight isometries with the sixteen grid origins at $s=2$, giving $K=128$ passes.

Phase-polarity reversal, illustrated separately in Figure~\ref{fig:phase-transformations}(b), replaces $\psi$ by $\mathcal W(-\psi)$ without moving the pixels. It is another way to vary the input; the sign of the result must then be reversed before alignment and averaging. This option is not used in the main 128-pass configuration.

Each pass is aligned by subtracting its minimum over the working image, then rounded to the encoded phase grid described in Appendix~\ref{app:implementation}. Let $\phi_k$ denote the aligned pass. The aligned passes are combined with the residual weights defined below. Averaging over a finite set of grid origins, spaced $s$ pixels apart, reduces the effect of the grid but does not make the method exactly invariant to translation.

\subsection{Residual-weighted averaging}
The weight of each pass comes from a residual score. The score uses the single-cycle operator
\begin{equation}
 S(t)=t-2\pi\mathbf1\{t>\pi\}+2\pi\mathbf1\{t<-\pi\}.
 \label{eq:single-reduction}
\end{equation}
which removes one cycle from a difference larger than $\pi$ in absolute value. Unlike the wrapping operator $\W$, it removes at most one cycle: a jump larger than $3\pi$ is not brought back into $[-\pi,\pi]$, and the values $\pm\pi$ are left unchanged. This matters for reconstructed passes, whose neighbor differences can be large, unlike those of the input. The residual score is
\begin{equation}
 L_S f=\operatorname{div}(S(\nabla f)),\qquad
 R_k^S=\frac{1}{P}\sum_{p}|L_S\phi_k(p)-L_S\psi(p)|,
 \label{eq:residual}
\end{equation}
where $P$ is the number of pixels of the padded working image, and both fields use the same boundary convention. The score is thus a mean over pixels. Because $S$ removes full-cycle jumps, it differs from the ordinary Poisson residual $L\phi_k-\rho$: it does not see an error of one full cycle. With $R_k=R_k^S$, the weights are
\begin{equation}
 \alpha_k=\frac{\exp[-(R_k-R_{\min})/\tau]}
 {\sum_j\exp[-(R_j-R_{\min})/\tau]},\quad
 \widehat\phi=\sum_k\alpha_k\phi_k,
 \label{eq:weighted}
\end{equation}
with $R_{\min}=\min_jR_j$. The parameter $\tau$ sets how strongly the weights favor passes with a small residual; we use $\tau=3(2\pi/256)$ radians, that is, three encoded phase levels. Subtracting $R_{\min}$ does not change the weights; it only prevents all the exponentials from underflowing to zero.

The residual measures how well the Laplacian of a pass agrees with that of the observation. It does not guarantee the correct number of cycles. The evaluation therefore judges the final average with reference-based phase errors and error maps.

\input{results_computation.tex}

\section{Reconstruction Accuracy and Observation Congruence}
\label{sec:evaluation-measures}
Two distinct questions guide the evaluation: how closely a reconstruction agrees with an available unwrapped reference, and how faithfully it reproduces the wrapped observation. Reference-based errors address the first question. The proposed corrected circular rewrapping error addresses the second, including when no unwrapped reference is available.

\subsection{Reference-based errors and diagnostic quantities}
Let $\Omega$ be the valid evaluation mask and $\phi_{\rm ref}$ the available reference, which need not equal the underlying physical phase. For an estimate $\widehat\phi$, define
\begin{equation}
 e(p)=\widehat\phi(p)-\phi_{\rm ref}(p)-\frac{1}{|\Omega|}\sum_{q\in\Omega}(\widehat\phi(q)-\phi_{\rm ref}(q)).
\end{equation}
We report root-mean-square error (RMSE) $=(|\Omega|^{-1}\sum_\Omega e^2)^{1/2}$, mean absolute error (MAE) $=|\Omega|^{-1}\sum_\Omega|e|$, the 95th percentile of $|e|$, and $C(t)=|\Omega|^{-1}\sum_\Omega\mathbf1\{|e|<t\}$. $C(\pi)$ is the fraction of pixels with an error below $\pi$; it is not an exact count of pixels with the correct number of cycles. Subtracting the mean difference removes the arbitrary global phase offset, and we use this mean alignment throughout. It does not change phase amplitudes or errors that vary in space. We also report the familiar error fraction $F_{>\pi}=|\Omega|^{-1}\sum_\Omega\mathbf1\{|e|>\pi\}$. At exact threshold ties, $1-C(\pi)$ and $F_{>\pi}$ differ because the former includes $|e|=\pi$.

The structural similarity index (SSIM) measures local structural agreement \cite{wang2004image}. We use two explicitly defined block-based versions. For contrast-normalized evaluation, SSIM$_{\rm mm}$ uses separate min--max normalization of the two full rectangular images to $[0,255]$, nonoverlapping $8\times8$ blocks, sample variances and covariance with denominator 63, and constants $(0.01L)^2$ and $(0.03L)^2$ with $L=256$. This score intentionally measures structure after contrast normalization.

SSIM$_{\rm add}$ is sensitive to amplitude. The estimate is first aligned by the mean offset; the minimum of the valid reference is then subtracted from both fields, which gives them a common origin without changing their amplitude. Here $L$ is the range of the valid reference phase, and only blocks entirely inside the evaluation mask are used. The two SSIM conventions are named separately throughout. They complement reference-based phase errors rather than replace them.

Gradient RMSE is the RMSE of forward differences, computed over the pixel pairs whose two pixels lie in $\Omega$. Lower values indicate closer agreement with the reference in local phase changes. A constant regional offset affects this score only at the region boundary, so it complements rather than replaces phase RMSE and MAE.

The amplitude slope
\begin{equation}
 a=\frac{\operatorname{cov}_{\Omega}(\phi_{\rm ref},\widehat\phi)}{\operatorname{var}_{\Omega}(\phi_{\rm ref})}
\end{equation}
is defined when the reference has nonzero variance and measures the linear amplitude scaling relative to that reference. The ideal slope is $a=1$; for example, $a=0.8$ indicates a fitted amplitude contraction of 20\%. The dimensionless diagnostic $|a-1|$ measures this departure for each image. For the gain-1.2 control in Figure~\ref{fig:metrics}, $a=1.2$ and $|a-1|=0.2$. This absolute error does not distinguish contraction from amplification and does not summarize spatial reconstruction errors.

As an exploratory regional diagnostic, we also report the size of the largest four-connected component of $\{p\in\Omega:|e(p)|\ge\pi\}$ divided by $|\Omega|$. Its meaning depends on the chosen threshold, gauge, and spatial resolution.

\subsection{Corrected circular rewrapping error}
\label{sec:maeuw2}
Congruence means that rewrapping a reconstruction reproduces the observed phase, after allowing for a global phase offset. An unconstrained least-squares solution can depart from this condition because it fits gradients rather than individual phase values. The proposed congruence measure quantifies this departure using only the reconstruction $\widehat\phi$ and the observation $\psi$.

Rewrapping the reconstruction and taking an ordinary absolute difference is insufficient: values $-\pi+\epsilon$ and $\pi-\epsilon$ differ by $2\pi-2\epsilon$ on the line but only $2\epsilon$ on the circle. We therefore wrap the difference as well. To remove the arbitrary global phase origin, define
\begin{equation}
 \begin{split}
 z(p)&=\W(\widehat\phi(p)-\psi(p)),\\
 m&=\frac{1}{|\Omega|}\sum_{p\in\Omega}e^{\mathrm i z(p)},\qquad
 \beta=\arg m.
 \end{split}
\end{equation}
The circular mean $\beta$ supplies one global offset \cite{mardia2000directional}. We define the corrected circular rewrapping error as the mean absolute angular discrepancy after this alignment:
\begin{equation}
 \begin{split}
 \mathrm{MAE}_{\mathrm{uw2}}
 &=\frac{1}{|\Omega|}\sum_{p\in\Omega}
   |\W(\W(\widehat\phi(p)-\beta)-\psi(p))|\\
 &=\frac{1}{|\Omega|}\sum_{p\in\Omega}|\W(z(p)-\beta)|.
 \end{split}
 \label{eq:congruence}
\end{equation}
The subscript $\mathrm{uw2}$ denotes the two wrapping operations. The inner wrap returns the reconstruction to the observation's phase interval; the outer wrap measures the shortest angular discrepancy. For a fixed $\beta$, each summand also equals
\begin{equation}
 \min_{k\in\mathbb Z}|\widehat\phi(p)-\beta-\psi(p)-2\pi k|.
\end{equation}
The score, in radians, is therefore the mean distance from each reconstructed value to the nearest value compatible with the observation. For $|m|>0$, it does not change when a global constant, or an integer multiple of $2\pi$ at any pixel, is added to the reconstruction. The circular mean removes a common offset, but it is not the offset that minimizes this $L_1$ score. When $|m|$ is close to zero, $\beta$ is unstable; at $m=0$ it is undefined, and our convention $\beta=0$ is arbitrary.

$\mathrm{MAE}_{\mathrm{uw2}}$ compares how faithful different methods are to the observation, without an unwrapped reference. The double wrap avoids artificial discrepancies at the phase-interval boundary, and the circular alignment removes an arbitrary global offset. This is useful for assessing how much a least-squares reconstruction departs from the measured phase, including on the experimental images in Figure~\ref{fig:real-gallery}.

\begin{figure*}[!t]\centering
\includegraphics[width=\textwidth]{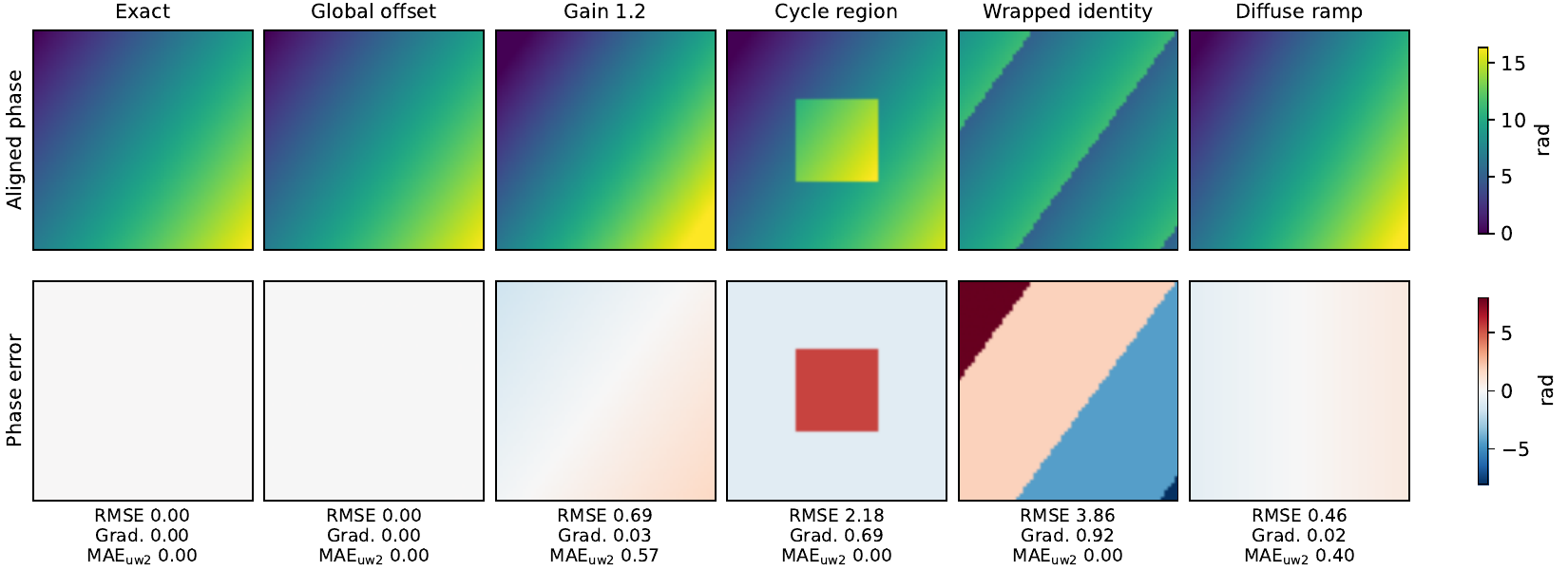}
\caption{Complementary responses of phase-error metrics and the proposed $\mathrm{MAE}_{\mathrm{uw2}}$ on a synthetic ramp. Reference-based errors use mean additive alignment; $\mathrm{MAE}_{\mathrm{uw2}}$ uses the circular offset in \eqref{eq:congruence}. A global offset leaves reference errors unchanged; amplitude scaling and a spatial cycle error remain visible in reference-based metrics. Both a cycle-shifted region and the wrapped identity can have zero $\mathrm{MAE}_{\mathrm{uw2}}$. Phase and error color scales are shared within their respective rows.}\label{fig:metrics}
\end{figure*}

$\mathrm{MAE}_{\mathrm{uw2}}$ measures consistency with the observation, not overall accuracy. For a noiseless observation $\psi=\W(\phi)$, it is also an error relative to the true phase modulo $2\pi$, after offset alignment. With noise, a reconstruction that preserves the noisy observation can score better than one that estimates the noise-free phase: denoising can increase the score. Moreover, both the wrapped identity $\widehat\phi=\psi$ and a region displaced by $2\pi$ can score zero, as shown in Figure~\ref{fig:metrics}; the metric therefore cannot establish correct cycle recovery. Its spatial mean also does not locate errors, and a small value can conceal localized discrepancies. We therefore read it together with reference-based phase errors, threshold fractions, and error maps whenever a reference is available, and never as a ranking of reconstruction quality on its own.

\section{Experimental Protocol}
\label{sec:protocol}
We use the measures defined in Section~\ref{sec:evaluation-measures} to compare all methods under a common phase-only input protocol. This section specifies the images, reconstruction settings, and accuracy and timing summaries.

\subsection{Images and reference phases}
The experiments use 202 distinct input--reference pairs spanning interferometric synthetic aperture radar (InSAR), magnetic resonance imaging (MRI), holography, synthetic phase fields, and intensity-derived test images. Exact duplicate pairs are removed before aggregation. The references have three origins: phases known by construction, physical-model estimates derived from elevation data, and reconstructions obtained with another unwrapper. With the last two, the scores measure agreement with a reference that can itself contain errors. Intensity-derived images provide controlled textures and discontinuities, including the Tree example in Figure~\ref{fig:tree-regions}; they are not simulations of an acquisition process.

The reference-bearing MRI images are slices and echoes of the Quantitative Susceptibility Mapping (QSM) Challenge 2.0 simulated head \cite{marques2021qsm}. The experimental holographic references were generated by quality-guided phase unwrapping (QGPU) in the dataset of Gontarz et al.\ \cite{gontarz2023phase}. The real InSAR examples come from InSAR-DLPU, whose reference phases are derived from elevation data \cite{zhou2024insardlpu}. Figure~\ref{fig:modality-gallery} illustrates these three reference types. Images sharing a reference are grouped for the uncertainty estimates in Section~\ref{sec:timing}.

\subsection{Compared methods and reconstruction settings}
The proposed method, labeled Tiled-DCT $8\times8$ (ours) in the figures and tables, uses 128 passes, with the joining and residual-weighting parameters fixed across all images. We compare it with global DCT least squares, SRNCP, SNAPHU, ROMEO, Delaunay $L^1$, DigiHolo, and a Goldstein-style branch-cut implementation. Global DCT uses a Neumann solve in floating point; SRNCP uses scikit-image with seed 42 for each image; SNAPHU 2.0.7 uses its smooth mode with unit input amplitude.

DigiHolo uses the binary of Antonopoulos et al.\ \cite{antonopoulos2015tilebased} with \texttt{polynom-2-3}, the SRNCP merger, and a target tile size of 16 pixels; its tile-count hint ensures at least two pixels per peripheral tile dimension. ROMEO uses the official mritools 4.7.1 binary on a single slice, with unit magnitude and an all-valid reconstruction mask; no multi-echo information is supplied. Delaunay $L^1$ uses the IPOL implementation with redundancy parameter $k=0$, minimum-cost flow, and reference vertex zero.

The Goldstein-style comparator is a Python reimplementation with greedy Euclidean pairing of opposite-sign residues and path integration avoiding cuts. Its pairing rule differs from the original growing-box construction; runtime and failure observations therefore concern this implementation.

The proposed C implementation uses the phase encoding in Appendix~\ref{app:implementation}; the comparators receive the original phase values. Evaluation masks are applied after reconstruction for every method.

\subsection{Timing, completion criteria, and statistical summaries}
\label{sec:timing}
Timing uses an Apple M3 Max with a budget of 12 numerical threads per implementation and one reconstruction running at a time. The proposed method distributes complete passes across 12 OpenMP workers and accumulates weighted contributions on the fly; implementations without parallel support remain serial. Each image--method pair receives one warmup followed by three measured calls. We take the median for each pair, then the median across the selected images. The timed call includes phase conversion and any subprocess startup, file transfer, and output decoding required by the implementation; dataset loading and metric evaluation are excluded. In particular, ROMEO timings include binary startup and NIfTI transfers, and Goldstein timings refer to the specified Python reimplementation. These times describe the latency of each implementation as used, not the complexity of the algorithms.

Each reconstruction has a 420\,s time limit. A case counts as complete only when the warmup and all three timed calls return finite outputs. A timeout, process error, or nonfinite output is recorded as a failure, without replacing missing phase values by zero. Completion counts accompany accuracy summaries. Methods that complete every image are compared on the full set. Comparisons that include an incomplete method use, for every method, the images that all methods completed. These comparisons can leave out difficult cases, so we report them in addition to the full-set results, not instead of them.

For paired comparisons, we report mean per-image differences and percentile 95\% confidence intervals from 10\,000 bootstrap resamples. Images sharing the same reference form one resampling group; each sampled group retains all its images, and the statistic is an image-weighted mean. These intervals describe variation among the evaluated images, rather than across independent subjects or acquisitions.

\section{Results}
\label{sec:results}
We first compare aggregate accuracy and completion, then examine the spatial errors that those summaries can hide. We then show reconstructions with and without references, and finally the measured speed--accuracy trade-offs.

\subsection{Quantitative comparison with reference methods}
\label{sec:baseline-results}
\begin{figure*}[!t]\centering
\includegraphics[width=0.93\textwidth]{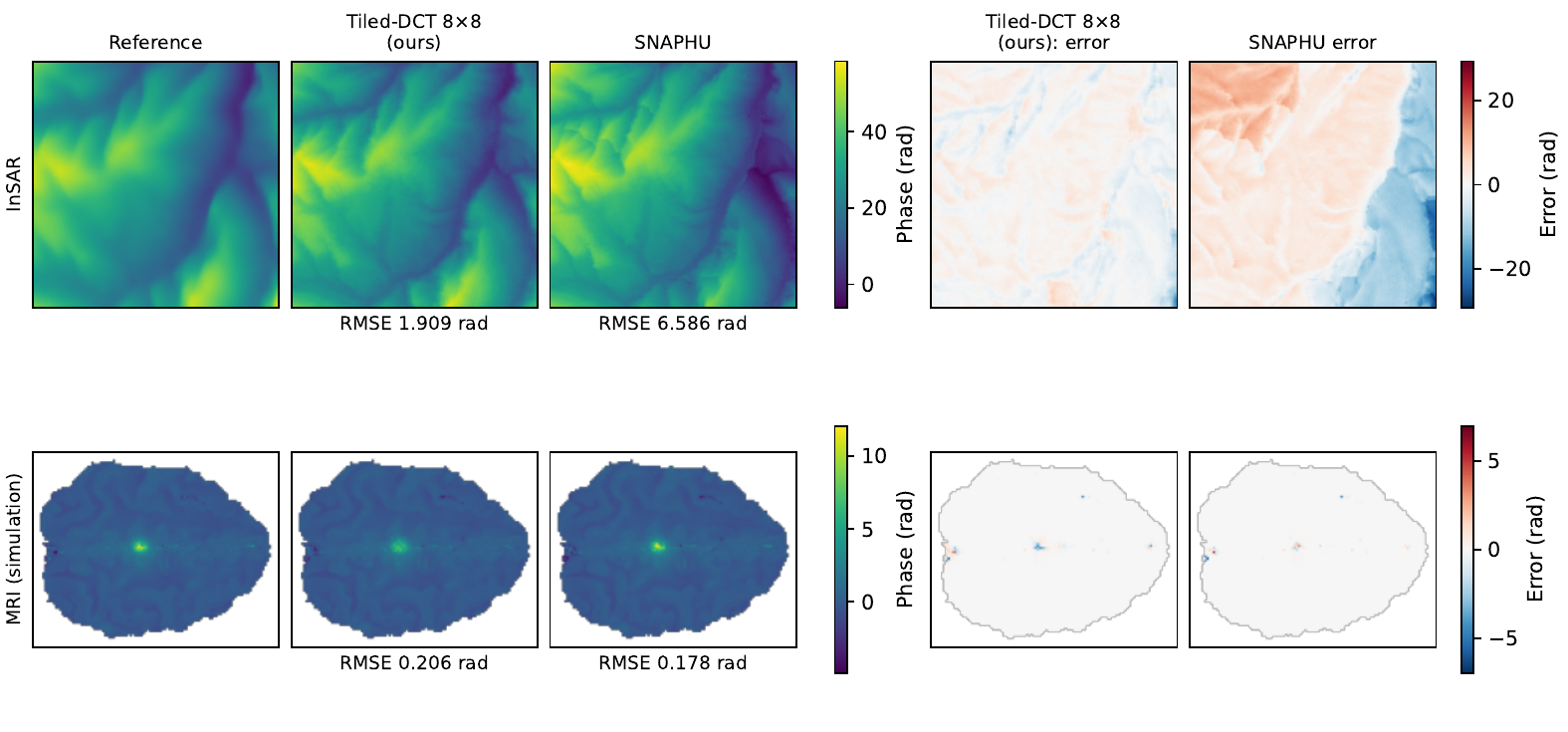}
\caption{Reconstructions and reference differences for InSAR (top) and an axial MRI simulation (bottom). The InSAR reference is a phase inferred from a digital elevation model, not an exact measurement of the true interferometric phase; its uncertainty and mismatch with the acquisition contribute to the displayed differences. The MRI reference is known within the QSM simulation model \cite{marques2021qsm}; it is not an independently measured in-vivo ground truth. The absence of exact phase truth in experimental acquisitions motivates distinguishing these reference types. Reconstructions share one phase scale per row after mean additive alignment, and signed differences share one symmetric scale per row. Pixels outside the evaluation mask are blank.}\label{fig:examples}
\end{figure*}

\input{tables/multimetric.tex}
\input{results_baselines.tex}
Figure~\ref{fig:examples} illustrates reconstruction on an InSAR image and an axial MRI slice. The InSAR example (top row) is the upper-middle case ranked by the proposed method's RMSE among the elevation-reference images. The simulated MRI slice (bottom row) was selected for its visible spatial phase variation; it illustrates local differences between reconstructions rather than an aggregate ranking.

\input{results_tree.tex}

\input{results_galleries.tex}

\input{results_pareto.tex}

\section{Discussion}
\label{sec:discussion}
The tile-based method combines three reconstruction operations: local gradient fitting inside tiles, spatial reconciliation across their boundaries, and residual-weighted averaging across shifted and transformed partitions. Table~\ref{tab:multimetric} evaluates this complete reconstruction against complementary reference methods. Its phase errors, threshold fractions, and structural similarity describe aggregate performance, while Figures~\ref{fig:examples} and~\ref{fig:tree-regions} locate the spatial discrepancies. What distinguishes the method is how these steps work together: the wrapped phase difference across each border in \eqref{eq:sample} separates the tile offset from the true phase variation, and the seam spread chooses which constraints are used before the passes are averaged.

The same design explains a remaining failure. With the propagation rule \eqref{eq:propagation}, one wrong seam in the tree shifts every tile behind it. A small MAD does not exclude a seam that is consistently wrong by one cycle. Neither the optimal tree of \eqref{eq:E0} nor a small residual guarantees the best phase recovery.

The proposed $\mathrm{MAE}_{\mathrm{uw2}}$ in \eqref{eq:congruence} separates observation fidelity from reference accuracy in Table~\ref{tab:multimetric}. A nonzero value quantifies mean departure from the observation modulo $2\pi$; a near-zero value indicates small average departure, without establishing correct cycles or excluding localized errors. Because denoising can increase this departure, a lower score does not by itself imply a more accurate reconstruction. The cycle-shift and amplitude-scaling controls in Figure~\ref{fig:metrics} demonstrate this distinction; the Tree error maps in Figure~\ref{fig:tree-regions} show its relevance to reconstruction. Figure~\ref{fig:real-gallery} applies the congruence measure to images without an unwrapped reference.

The speed--accuracy fronts in Figure~\ref{fig:pareto} show how the preferred method changes with the error criterion and its aggregation. This comparison includes every implementation on identical successful cases, while Table~\ref{tab:multimetric} retains their completion counts and the full-set accuracy of complete methods. Both views are needed: restricting a front to successful reconstructions can remove precisely the difficult inputs that matter in deployment. The reported latency also includes implementation-specific startup and data conversion. Figures~\ref{fig:synthetic-gallery}--\ref{fig:real-gallery} complement these aggregate trade-offs by locating regional reconstruction differences.

The independent passes also provide a natural unit of parallel work. The accumulation scheme in Section~\ref{sec:computation} reduces pass-image storage from $O(KP)$ to $O(n_{\rm th}P)$, making it possible to retain the diversity of shifted grids and image isometries without storing all reconstructed images. The optimized DCT implementation connects the scaled split-radix factorization of Shao and Johnson \cite{shao2008typeii} with the SIMD implementation work described in BRiDCT \cite{moevus2026bridct}. A lower arithmetic count does not by itself guarantee a shorter execution time: data movement, vectorization, tile joining, and residual evaluation also contribute to the measured latency. The speed--accuracy comparison in Figure~\ref{fig:pareto} therefore uses the complete reconstruction with the selected kernels, rather than extrapolating from transform timings.

Adaptive tiling is another possible way to reduce reconstruction work. In our study of quadtree and kd-tree partitions \cite{moevus2026adaptive}, the tested optimized adaptive configurations used fewer tiles but remained slower than the optimized regular grid in single-threaded experiments. Partition construction and larger local solves outweighed the savings at tile boundaries. This result supports retaining a regular grid here; it does not establish a universal disadvantage of adaptive tiling. A useful future direction is to compare complete runtime at a prescribed reconstruction accuracy, including limited refinement and parallel execution.

The experiments cover several image types, but do not establish generalization to independent acquisitions. The reference types defined in Section~\ref{sec:protocol} and illustrated in Figures~\ref{fig:examples} and~\ref{fig:modality-gallery} retain their own uncertainties when inferred from physical models or reconstruction algorithms. The phase-only comparisons also leave aside acquisition-specific information such as magnitude or coherence, while masks are used for evaluation rather than reconstruction. These conditions limit how far the reported accuracy can be generalized.

\section{Conclusion}
We have presented a tile-based phase-unwrapping method combining local DCT least-squares reconstruction, reliability-ordered tile joining, and residual-weighted averaging over shifted grids and image isometries. Parallel execution, with each pass added to the average as soon as it is computed, makes it possible to average many passes without storing them. With $8\times8$ tiles and 128 passes, the method achieves the lowest mean fraction of errors above $\pi$ among methods completing the evaluation (Table~\ref{tab:multimetric}), while other methods attain lower median RMSE. It lies on several speed--accuracy fronts in Figure~\ref{fig:pareto}; the preferred method therefore depends on the metric and the required runtime. The proposed corrected circular rewrapping error quantifies fidelity to the observed phase modulo $2\pi$ without an unwrapped reference. The controls in Figure~\ref{fig:metrics} and the Tree reconstructions in Figure~\ref{fig:tree-regions} show why observation fidelity, regional cycle errors, and diffuse phase distortion require complementary evaluation. Joining the tiles in order of reliability and averaging many passes reconcile the local solutions while varying how errors propagate. Evaluating new acquisitions and adaptive partitions at a prescribed reconstruction accuracy would clarify the range of conditions in which this combination is most useful.

\bibliographystyle{IEEEtran}
\bibliography{refs}

\appendices
\section{Numerical Implementation Details}
\label{app:implementation}
The configuration uses tile side $\NT=8$, grid stride $s=2$, pass count $K=128$, residual parameter $\tau=3q$, seam constant $c=q$, and spectral threshold $T_n=0$, where $q=2\pi/256$ radians per encoded phase unit. All eight square isometries are included.

Inputs are encoded by rounding $(\psi+\pi)256/(2\pi)$ modulo 256. An input of $M$ rows and $N$ columns is padded with zeros to a square of side $\max(M,N)$, and the output is cropped back to its original shape. Local tiles extending outside the working square use nearest-border replication. The residual in \eqref{eq:residual} is evaluated over the full working square.

For local differences and seams, the single-cycle helper retains both $-\pi$ and $+\pi$ at exact ties. The local C output uses the conversion \texttt{(int)(x+0.5)}, which truncates toward zero after adding one half. After a pass is returned to the input frame, its minimum is subtracted and nonnegative values are rounded to the same grid. Each OpenMP worker maintains a double-precision weighted accumulator and rescales it when a smaller residual is encountered. The final reduction brings these private accumulators to a common residual minimum before summation, implementing \eqref{eq:weighted} without storing every pass. The evaluated parallel configuration uses 12 workers.

For contrast-normalized SSIM, constant images are mapped to zero instead of dividing by a zero range. Images smaller than one full block have no SSIM value.

\clearpage
\onecolumn
\input{appendix_galleries.tex}
\end{document}

%% file: results_computation.tex
\subsection{Parallel implementation and accumulation on the fly}
\label{sec:computation}
A sequential implementation stores all $K$ passes before averaging them. Ours instead gives each complete pass to one worker. The worker solves and joins the pass, maps it back to the input frame, computes its residual, and immediately adds it to a private weighted sum. Only these sums are combined at the end. The average is thus computed on the fly, without storing or rereading the $K$ reconstructions. For $P$ working pixels and $n_{\rm th}$ workers, the memory used for pass images drops from $O(KP)$ to $O(n_{\rm th}P)$, plus shared buffers for the transforms and the joining. Precomputed spectral denominators and contiguous memory access further reduce repeated work. The weights use the stable normalization in \eqref{eq:weighted}.

The evaluated configuration uses our optimized fixed-size single-instruction, multiple-data (SIMD) DCT solver with $8\times8$ tiles; a corresponding routine supports $16\times16$ tiles. Its separable DCT-II and DCT-III routines follow the scaled split-radix algorithm of Shao and Johnson \cite{shao2008typeii}. Our BRiDCT work \cite{moevus2026bridct} describes the associated implementation approach, including SIMD organization, register blocking, and numerical verification. The inverse follows the transposed computation with the appropriate normalization. Four independent one-dimensional transforms are evaluated together using SIMD operations, and blocked transpositions connect the two spatial axes. These routines only speed up the local least-squares solve in \eqref{eq:dct}; they do not change the method.

%% file: tables/multimetric.tex
\begin{table*}[!t]\centering\footnotesize
\setlength{\tabcolsep}{4pt}
\caption{Full-set reconstruction accuracy and completion counts ($n$). A dash denotes an unavailable full-set aggregate, rather than zero error; successful-case results for all methods are reported separately in Table~\ref{tab:common-success}. Phase errors and MAE$_{\rm uw2}$ are in radians. $F_{>\pi}$ is the mean percentage of pixels above $\pi$; other scores are per-image medians. SSIM$_{\rm mm}$ uses contrast normalization and SSIM$_{\rm add}$ preserves amplitude. The proposed method uses $8\times8$ tiles and 128 passes.}\label{tab:multimetric}
\begin{tabular}{@{}lrrrrrrrr@{}}\toprule
Method & $n$ & MAE & RMSE & Grad. RMSE & $F_{>\pi}$ & SSIM$_{\rm mm}$ & SSIM$_{\rm add}$ & MAE$_{\rm uw2}$\\\midrule
Global DCT & 202 & 0.802 & 1.052 & 0.640 & 20.547 & 0.748 & 0.830 & 0.468\\
SRNCP & 202 & 0.527 & 0.816 & 0.783 & 14.319 & 0.721 & 0.859 & 0.000\\
SNAPHU & 202 & 0.474 & 0.729 & 0.702 & 13.930 & 0.783 & 0.880 & 0.000\\
ROMEO & 202 & 0.505 & 0.726 & 0.771 & 15.967 & 0.736 & 0.847 & 0.000\\
Delaunay L1 & 202 & 0.521 & 0.824 & 0.750 & 16.482 & 0.752 & 0.850 & 0.000\\
DigiHolo & 202 & 0.588 & 0.952 & 0.802 & 18.359 & 0.728 & 0.818 & 0.000\\
Goldstein (reimpl.) & 192 & -- & -- & -- & -- & -- & -- & --\\
\textbf{Tiled-DCT $8\times8$ (ours)} & 202 & 0.580 & 0.855 & 0.655 & 12.656 & 0.769 & 0.874 & 0.170\\
\bottomrule\end{tabular}\end{table*}

\begin{table*}[!t]\centering\footnotesize
\setlength{\tabcolsep}{4pt}
\caption{Accuracy on the 192 images for which all eight implementations completed reconstruction. The column $n$ gives the number of images. Every row uses the same images; this conditional comparison omits cases on which any implementation failed. Phase errors and MAE$_{\rm uw2}$ are in radians. $F_{>\pi}$ is the mean percentage of pixels above $\pi$; other scores are per-image medians. SSIM$_{\rm mm}$ uses contrast normalization and SSIM$_{\rm add}$ preserves amplitude. The proposed method uses $8\times8$ tiles and 128 passes.}\label{tab:common-success}
\begin{tabular}{@{}lrrrrrrrr@{}}\toprule
Method & $n$ & MAE & RMSE & Grad. RMSE & $F_{>\pi}$ & SSIM$_{\rm mm}$ & SSIM$_{\rm add}$ & MAE$_{\rm uw2}$\\\midrule
Global DCT & 192 & 0.767 & 1.003 & 0.615 & 19.556 & 0.754 & 0.834 & 0.396\\
SRNCP & 192 & 0.491 & 0.755 & 0.761 & 12.926 & 0.740 & 0.865 & 0.000\\
SNAPHU & 192 & 0.458 & 0.714 & 0.672 & 14.062 & 0.783 & 0.880 & 0.000\\
ROMEO & 192 & 0.481 & 0.703 & 0.740 & 14.557 & 0.755 & 0.849 & 0.000\\
Delaunay L1 & 192 & 0.505 & 0.763 & 0.716 & 16.065 & 0.758 & 0.855 & 0.000\\
DigiHolo & 192 & 0.570 & 0.889 & 0.784 & 17.137 & 0.734 & 0.828 & 0.000\\
Goldstein (reimpl.) & 192 & 0.821 & 1.308 & 0.800 & 23.637 & 0.718 & 0.791 & 0.000\\
\textbf{Tiled-DCT $8\times8$ (ours)} & 192 & 0.572 & 0.805 & 0.635 & 11.752 & 0.779 & 0.880 & 0.149\\
\bottomrule\end{tabular}\end{table*}

%% file: results_baselines.tex
On the full test set, the proposed method achieves median RMSE 0.855 radians, mean $C(\pi)=0.873$, and median SSIM$_{\rm mm}=0.769$ (Table~\ref{tab:multimetric}). Among methods completing the full set, ROMEO has the lowest median RMSE (0.726 radians), SNAPHU the highest median SSIM$_{\rm mm}$ (0.783), and Tiled-DCT $8\times8$ (ours) the lowest mean fraction above $\pi$ (12.656\%). These summaries measure different aspects of reconstruction quality.

The paired mean RMSE contrasts (proposed minus comparator, in radians, with 95\% intervals) are SNAPHU: +0.021 [-0.193, +0.234]; SRNCP: -0.369 [-0.712, -0.051]; DigiHolo: -1.157 [-1.646, -0.762]; ROMEO: -0.394 [-0.622, -0.191]; Delaunay L1: -0.284 [-0.510, -0.067]. Negative contrasts favor the proposed method. Complete methods use the full set; any incomplete comparator uses the common successful cases. These image-group intervals describe the evaluated collection, not independent acquisition populations.

Goldstein (reimpl.) completed 192 of 202 cases, with one nonfinite output and nine timeouts. Table~\ref{tab:common-success} compares every method on the same 192 successful cases. Restricting to these cases can remove difficult phase fields; it is therefore a conditional comparison, not a replacement for the full-set results in Table~\ref{tab:multimetric}.

The congruence column in Table~\ref{tab:multimetric} provides a complementary diagnostic. Median MAE$_{\rm uw2}$ is 0.468 radians for global DCT and 0.170 radians for tiled DCT; it rounds to zero for SRNCP, SNAPHU, ROMEO, Delaunay L1, and DigiHolo. The near-zero values indicate consistency with the wrapped observation, not recovery of the correct cycles: their reference errors remain nonzero. Tiling thus brings the result closer to the observation than global DCT does, without forcing congruence at the end.

%% file: results_tree.tex
\subsection{Isolated cycle errors and diffuse phase errors}

\label{sec:tree-example}

\begin{figure*}[!t]\centering\includegraphics[width=0.94\textwidth]{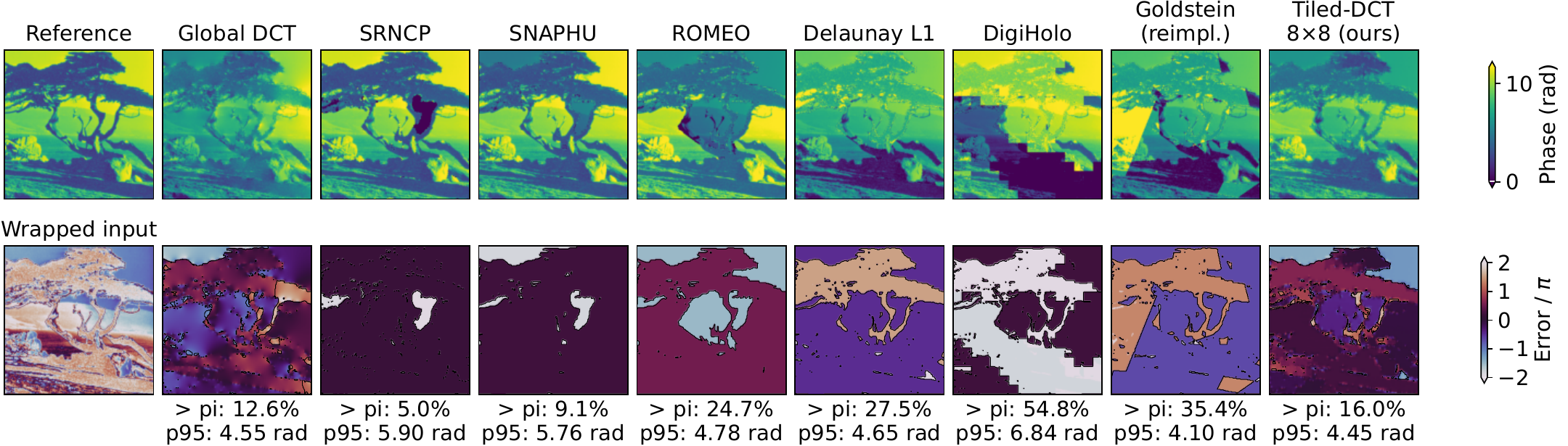}

\caption{Spatial error patterns on the Tree case, with seven reference algorithms and the proposed $8\times8$, 128-pass reconstruction. Top: exact constructed reference and reconstructions after mean-gauge alignment, on one common phase scale. Bottom: wrapped input, followed by signed phase errors on one symmetric cyclic color scale (matching colors at $-2\pi$ and $+2\pi$), with a black boundary around $|e|>\pi$. The final column shows Tiled-DCT $8\times8$ (ours), including residual weighting. Phase maps use the reference range; error colors saturate outside $[-2\pi,2\pi]$, marked by the colorbar extensions. All scores use unclipped values. Phase maps are not individually contrast-normalized. The regional errors of the congruent reconstructions and the diffuse least-squares errors motivate complementary metrics.}\label{fig:tree-regions}\end{figure*}

Goodhart's law cautions against treating success on an evaluation measure as success at the underlying task \cite{strathern1997improving}. For phase unwrapping, favorable aggregate scores or exact congruence do not by themselves establish spatially correct reconstruction. Figure~\ref{fig:tree-regions} illustrates the distinction through isolated cycle errors, motivating the joint examination of numerical scores and spatial error maps.

Figure~\ref{fig:tree-regions} contrasts localized cycle errors with errors spread over the image, on the Tree image. The image is an intensity-derived stress test (USC-SIPI miscellaneous image 4.1.06, $256\times256$), with an exact constructed phase spanning 1.85 cycles. The top row compares the reconstructed phases, and the bottom row shows their signed errors. All errors use the same mean additive gauge as the quantitative tables.

In the SNAPHU and SRNCP columns of Figure~\ref{fig:tree-regions} (bottom row), the fractions of pixels beyond $\pi$ are 9.1\% and 5.0\%; their largest connected error regions occupy 5.0\% and 2.4\% of the image. Their $\mathrm{MAE}_{\mathrm{uw2}}$ values are below $10^{-5}$ radians: rewrapping does not expose the cycle displacement. With median alignment, their 95th-percentile errors are 6.28 and 6.28 radians, close to one full cycle. Mean alignment redistributes the global offset, explaining the slightly different tail values in Figure~\ref{fig:tree-regions}.

The additional reference methods in Figure~\ref{fig:tree-regions} also exhibit regional errors: the fractions beyond $\pi$ are ROMEO (24.7\%), Delaunay L1 (27.5\%), Goldstein (reimpl.) (35.4\%). Their MAE$_{\rm uw2}$ values are below $10^{-5}$ radians despite these errors. Congruence therefore does not distinguish their different recovered cycle fields.

The final column of Figure~\ref{fig:tree-regions} gives the proposed reconstruction, with a mean-gauge 95th-percentile error of 4.45 radians and a fraction beyond $\pi$ of 16.0\%. The bottom-row maps for global DCT and tiled DCT show a more diffuse spatial pattern than the regional displacements of SNAPHU and SRNCP. This example motivates reporting regional structure alongside MAE, RMSE, SSIM, and threshold fractions; which error pattern is worse depends on the application.

The cycle-shift control in Figure~\ref{fig:metrics} illustrates why a regional phase bias can remain fully consistent modulo $2\pi$. A diffuse sub-cycle error can also affect amplitudes and gradients. The tile architecture and reliability joining address how errors propagate, while the spatial maps show what remains after averaging. A largest-component statistic summarizes one aspect of that structure and complements the pixelwise metrics.

%% file: results_galleries.tex
\subsection{Reconstructions across textures and modalities}
The extended reconstruction galleries are collected in Appendix~\ref{app:galleries}. Figure~\ref{fig:synthetic-gallery} compares the methods on textured phase fields constructed from natural images. Within the Baboon (rows 1--2) and Lena (rows 3--4) pairs, the clean and noisy inputs share one reference, isolating the response to measurement perturbations. Other rows vary the texture or phase range. In Figure~\ref{fig:synthetic-gallery}, the noisy Lena row shows a large regional displacement for SRNCP, while the clean Lena row shows a tile-shaped discontinuity for DigiHolo. The global DCT and tiled DCT columns illustrate more spatially distributed changes, which the accompanying RMSE and threshold fractions summarize without locating.

Figure~\ref{fig:modality-gallery} extends the comparison to two InSAR scenes, two holographic acquisitions, and the simulated MRI slice of Figure~\ref{fig:examples}, now shown for all eight methods. The source datasets provide the experimental reference phases: the InSAR phases are derived from Shuttle Radar Topography Mission (SRTM) elevation data in InSAR-DLPU \cite{zhou2024insardlpu}, while the holographic phases are obtained by quality-guided phase unwrapping (QGPU) in the dataset of Gontarz et al.\ \cite{gontarz2023phase}. Agreement with these references helps compare spatial behavior but includes the assumptions and uncertainties of the reference itself. The MRI reference is known within the QSM simulation model \cite{marques2021qsm}.

\subsection{Observation congruence without an unwrapped reference}
Figure~\ref{fig:real-gallery} shows an experimental MRI head image and two holographic phase maps for which no unwrapped reference is available in the comparison. We therefore report $\mathrm{MAE}_{\mathrm{uw2}}$ rather than reference RMSE. In the MRI row of Figure~\ref{fig:real-gallery}, the displayed $\mathrm{MAE}_{\mathrm{uw2}}$ values round to zero for SRNCP, SNAPHU, ROMEO, Delaunay L1, DigiHolo, and Goldstein (reimpl.), despite their different spatial reconstructions. Global DCT and tiled DCT give 1.02 and 0.86\,rad, respectively. These values quantify observation fidelity through \eqref{eq:congruence}; they cannot establish which reconstruction is physically correct.

%% file: results_pareto.tex
\subsection{Speed--accuracy trade-offs}
\label{sec:pareto}
\begin{figure*}[!t]\centering
\includegraphics[width=.96\textwidth]{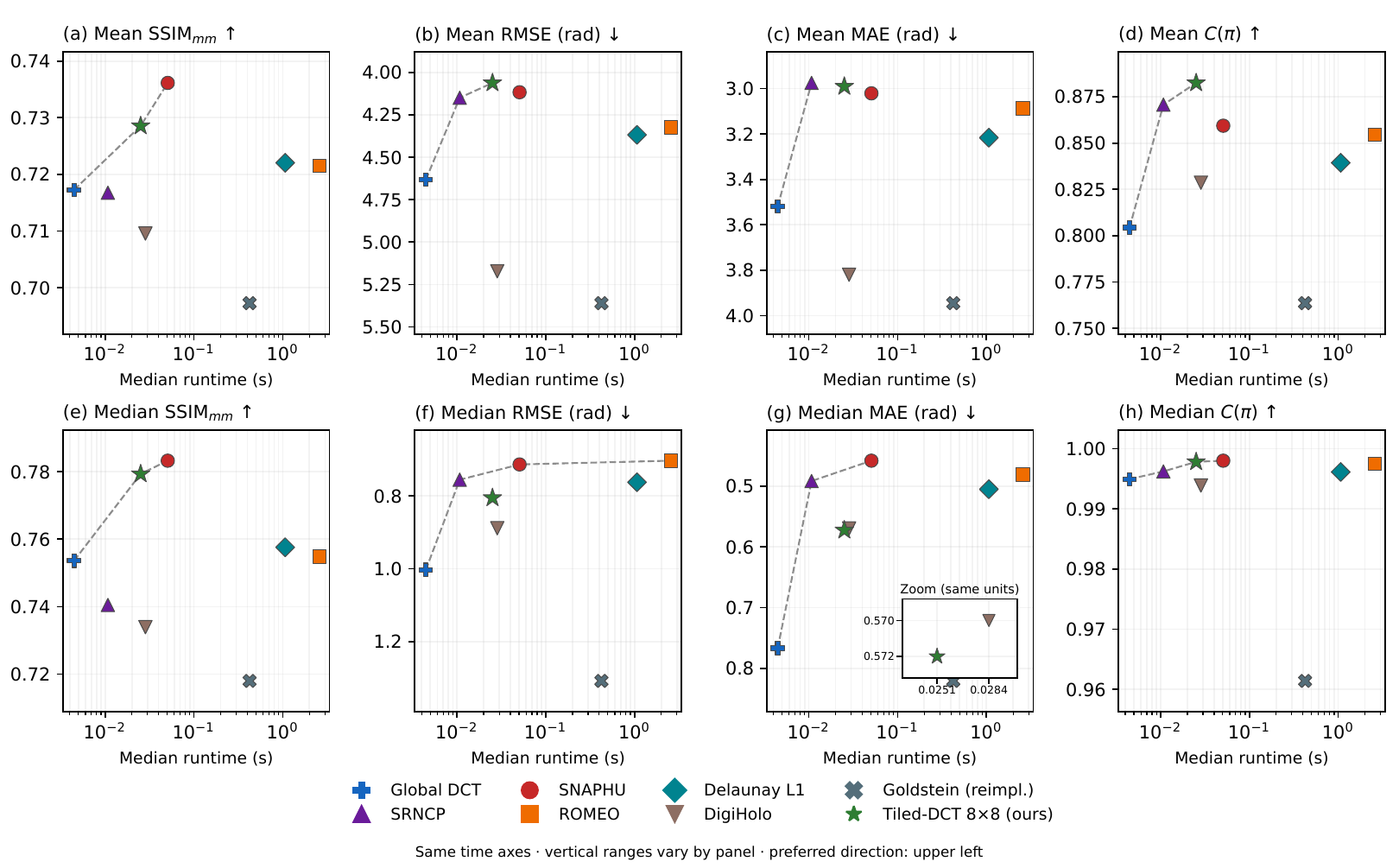}
\caption{Speed--accuracy trade-offs for eight implementations on the 192 common successful images of Table~\ref{tab:common-success}, with a budget of 12 numerical threads. Columns: SSIM$_{\rm mm}$, RMSE, MAE, and $C(\pi)$; rows: mean (top) and median (bottom) of per-image scores. Both coordinates use the same images for every method. Dashed lines connect nondominated point estimates, without uncertainty margins; lower-is-better axes are reversed, so the preferred direction is upper left. The logarithmic time axis is common; vertical ranges vary by panel. The inset in (g) enlarges the DigiHolo and tiled DCT points without changing their coordinates. Tiled DCT uses $8\times8$ tiles, 128 passes and 12 workers. These fronts are conditional on successful completion; Table~\ref{tab:multimetric} retains full-set completion and accuracy results.}\label{fig:pareto}
\end{figure*}

Using the timing protocol in Section~\ref{sec:timing}, we compare reconstruction time with each accuracy criterion. A Pareto front contains the tested methods for which no alternative is both no slower and no less accurate, with at least one strict improvement. Figure~\ref{fig:pareto} compares all eight implementations using the common successful images in Table~\ref{tab:common-success}. Both the accuracy and runtime summaries use that same support for every method. Columns show structural similarity, RMSE, MAE, and fraction below $\pi$; rows show their means and medians. MAE gives less emphasis to large phase errors than RMSE.

On this common support, tiled DCT belongs to the mean-score fronts for SSIM, RMSE, and $C(\pi)$ (Fig.~\ref{fig:pareto}(a), (b), and (d)) and to the median-score fronts for SSIM and $C(\pi)$ (panels (e) and (h)). Its median reconstruction time is 0.025\,s. The two rows distinguish average scores, which include the influence of large errors, from scores on a typical case. A front is relative to its error criterion and its included implementations; membership uses point estimates and does not establish statistical dominance. For median RMSE and MAE (panels (f) and (g)), SRNCP attains both a lower error and a shorter median runtime than tiled DCT.

The completion-conditioned front must be read with the full-set results in Table~\ref{tab:multimetric}. A method that fails on difficult inputs is not credited with zero error or an artificially short successful runtime on those inputs; its completion count remains a separate result. Excluding those inputs from every method makes the plotted comparison matched, but does not remove the selection effect.

%% file: appendix_galleries.tex
\section{Additional reconstruction galleries}
\label{app:galleries}
\begin{figure}[!htbp]\centering
\includegraphics[width=\textwidth,height=.83\textheight,keepaspectratio]{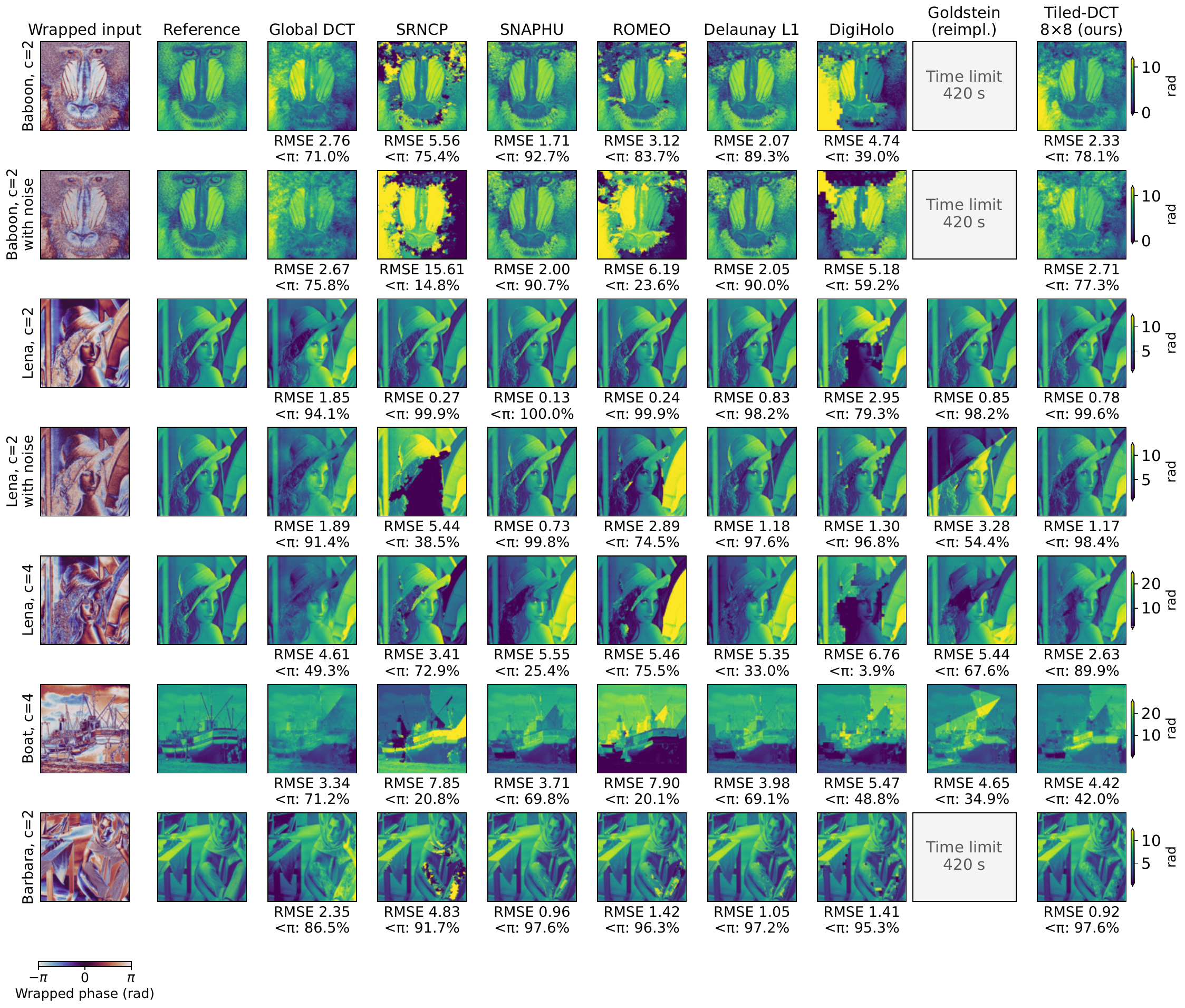}
\caption{Seven constructed phase fields and reconstructions by the eight methods. Columns: wrapped input, exact constructed reference, and reconstructions. The factor $c$ scales the intensity-derived phase range; the noisy Baboon and Lena inputs retain the corresponding clean reference. Every reference and reconstruction in a row shares the reference's phase scale after mean additive alignment. Colorbar extensions indicate values outside the display range; numerical scores use unclipped outputs. RMSE is in radians and $<\pi$ denotes the percentage of valid pixels below that error threshold. Wrapped inputs share the cyclic scale shown below their column. The proposed method uses $8\times8$ tiles and 128 passes. A labeled failure panel indicates that no finite reconstruction was available within the time limit; no accuracy score is assigned to it.}\label{fig:synthetic-gallery}
\end{figure}

\begin{figure}[!htbp]\centering
\includegraphics[width=\textwidth]{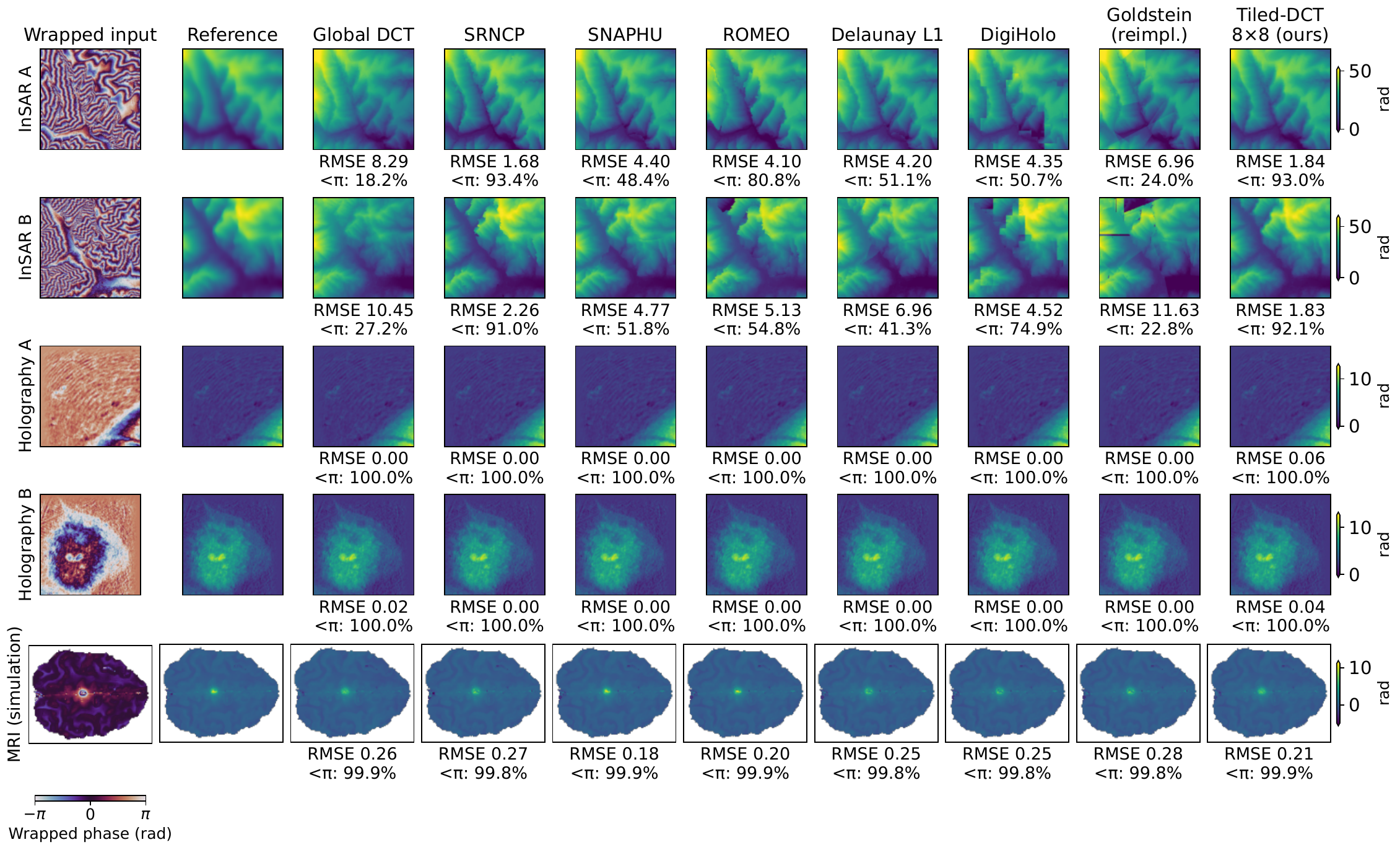}
\caption{Multimodal reconstruction using references supplied by the source datasets. Rows 1--2: real TanDEM-X interferograms from InSAR-DLPU; their reference phases are generated from Shuttle Radar Topography Mission (SRTM) elevation data \cite{zhou2024insardlpu}, using terrain information beyond the wrapped phase available to the compared methods. Rows 3--4: holographic acquisitions with references produced by quality-guided phase unwrapping (QGPU), selected by Gontarz et al.\ for its robustness in their holographic-tomography processing \cite{gontarz2023phase}. These InSAR and holographic references are comparison targets, not exact physical ground truths; their scores measure agreement with them. Row 5 shows an MRI slice with a phase known within the QSM simulation model \cite{marques2021qsm}. Columns and scores follow Figure~\ref{fig:synthetic-gallery}, with a common reference-based phase scale within each row.}\label{fig:modality-gallery}
\end{figure}

\begin{figure}[!htbp]\centering
\includegraphics[width=\textwidth]{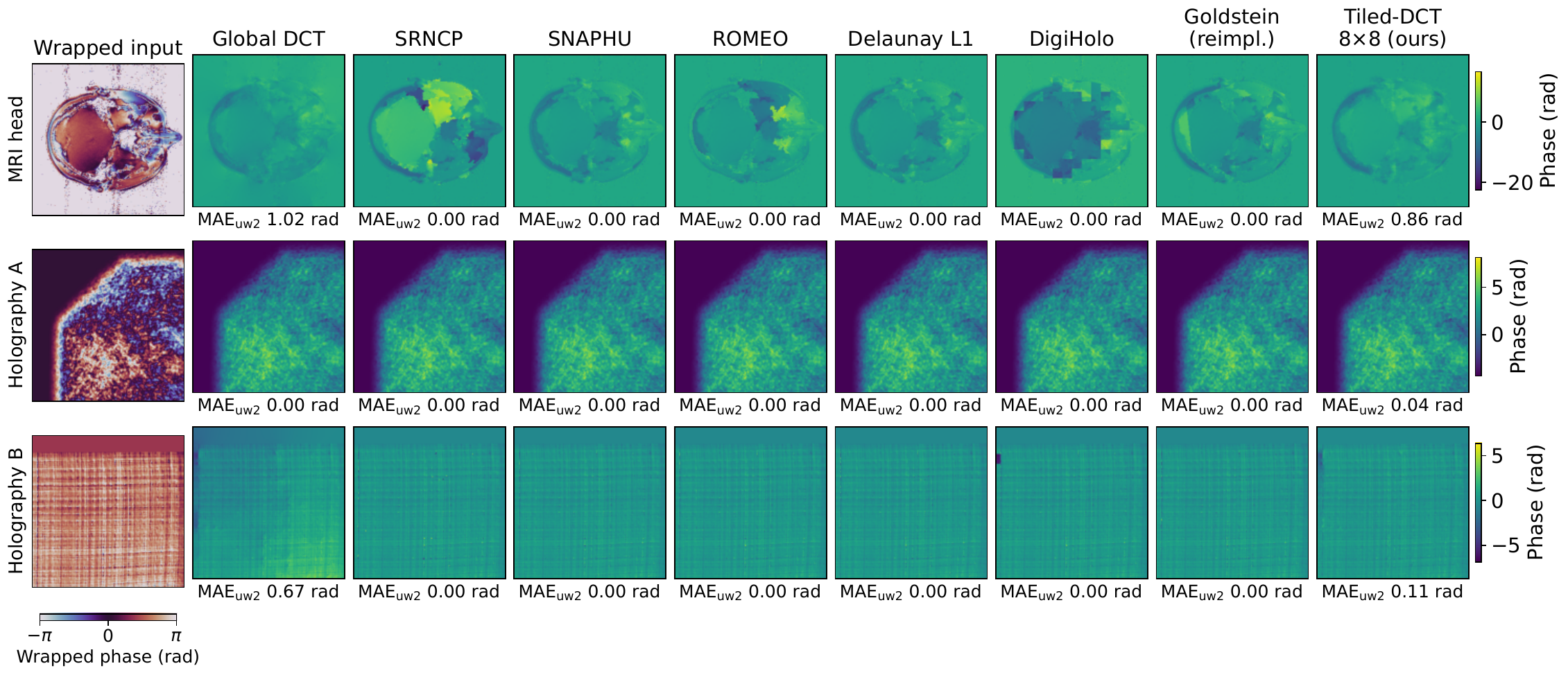}
\caption{Experimental phase maps without an unwrapped reference: an MRI head and two holographic examples. The first column is the wrapped observation; the remaining columns show the eight reconstructions. Each reconstruction is centered by its valid-pixel mean for display, and all reconstructed maps within a row share one phase scale. The reported $\mathrm{MAE}_{\mathrm{uw2}}$ uses circular offset alignment as defined in \eqref{eq:congruence}. No reference-error map or accuracy ranking is inferred from these images.}\label{fig:real-gallery}
\end{figure}

Figures~\ref{fig:synthetic-gallery}--\ref{fig:real-gallery} collect the extended comparisons discussed in Section~\ref{sec:results}. The three settings are constructed phase fields, multimodal images with dataset references, and experimental images without an unwrapped reference.

